\documentclass{ws-ijgmmp}

\begin{document}

\markboth{Zi-Hua Weng}
{Precessional angular velocity and field strength in the complex octonion space}

%%%%%%%%%%%%%%%%%%%%% Publisher's Area please ignore %%%%%%%%%%%%%%%
%
\catchline{}{}{}{}{}
%
%%%%%%%%%%%%%%%%%%%%%%%%%%%%%%%%%%%%%%%%%%%%%%%%%%%%%%%%%%%%%%%%%%%%

\title{Precessional angular velocity and field strength in the complex octonion space
%\footnote{For the title, try not to use more than 3 lines. Typeset in 10 pt roman, uppercase and boldface.}
}

\author{Zi-Hua Weng
%\footnote{Typeset names in 8 pt roman, uppercase. Use footnote to indicate permanent address of author.}
}
\address{School of Aerospace Engineering, Xiamen University, Xiamen, China
\\
College of Physical Science and Technology, Xiamen University, Xiamen, China
%\footnote{State completely without abbreviations, the affiliation and mailing address, including country. Typeset in 8 pt italic.}
\\
\email{xmuwzh@xmu.edu.cn
%\footnote{Typeset author's e-mail address in 8pt italic}
}
}

%\author{SECOND AUTHOR}
%\address{Group, Laboratory, Address\\ City, State ZIP/Zone, Country\\
%author\_id@domain\_name }

\maketitle

\begin{history}
\received{(Day Month Year)}
\revised{(Day Month Year)}
\end{history}

\begin{abstract}
  The paper aims to apply the octonions to explore the precessional angular velocities of several particles in the electromagnetic and gravitational fields. Some scholars utilize the octonions to research the electromagnetic and gravitational fields. One formula can be derived from the octonion torque, calculating the precessional angular velocity generated by the gyroscopic torque. When the octonion force is equal to zero, it is able to deduce the force equilibrium equation and precession equilibrium equation and so forth. From the force equilibrium equation, one can infer the angular velocity of revolution for the particles. Meanwhile, from the precession equilibrium equation, it is capable of ascertaining the precessional angular velocity induced by the torque derivative, including the angular velocity of Larmor precession. Especially, some ingredients of torque derivative are in direct proportion to the field strengths. The study reveals that the precessional angular velocity induced by the torque derivative is independent of that generated by the torque. The precessional angular velocity, induced by the torque derivative, is relevant to the torque derivative and the spatial dimension of precessional velocity. It will be of great benefit to understanding further the precessional angular velocity of the spin angular momentum.
\end{abstract}

\keywords{precessional angular velocity; quaternion; octonion; Larmor precession; torque; electromagnetic field; gravitational field.
\\
MSC[2010]: 83E15, 81R60, 17A35 }

\section{\label{sec:level1}Introduction}

In the magnetic fields, why the magnetic moment of a boson is distinct from that of a fermion? This problem has been perplexing and fascinating some scholars for a long time. They endeavor to account for this conundrum, attempting to reveal the discrepancy between the magnetic moment of the boson with that of the fermion, trying to deepen our understanding of the spin angular momentum. Until lately, the appearance of the octonionic electromagnetic and gravitational theories replies partially this puzzle. As one of theoretical applications, the associating inferences of this complex-octonion field theory can be applied to explore the magnetic moment and precessional angular velocity of charged particles. It is able to account for why the magnetic moment of a boson is different from that of a fermion in the magnetic fields, unpuzzling why the precessional motion always revolves around the external magnetic flux density for the charged particle.

Making use of the Foucault pendulum, the scholars are able to research the physical properties of the Earth's rotation, validating the rotation of the Earth. Some scholars discovered successively numerous physical phenomena relevant to the gyroscopic motions. According to the types of external physical quantities (Table \ref{tab:1}), the precessional motions, in the existing studies, can be separated into two categories, the precession generated by the torque, and the precession induced by the torque derivative (in Section 2). The precession induced by the torque (torque-induced precession, for short) meets the demand of the torque definition derived from the angular momentum (Appendix A). Meanwhile, the precession induced by the torque derivative obeys the precession equilibrium equation (in Section 4). Some ingredients of the torque derivative are in direct proportion to the external field strengths. And the precession induced by the external field strength (field-induced precession, for short) is a primary representative of the precession induced by the torque derivative.

The torque-induced precession can be applied to explore the axial precession and perihelion precession for the Earth and other planets. Condurache \emph{et al.} \cite{FP1} offered a comprehensive study of the motion in a central force field with respect to a rotating non-inertial reference frame. Stanovnik \cite{FP2} compared the precession of a Foucault pendulum to the precession of an ideal elastic pendulum, for which the string force is proportional to string length. Icaza-Herrera \emph{et al.} \cite{FP3} offered a Lagrangian approach in the case when the angular velocity is the sidereal angular velocity of the Earth. Bergmann \emph{et al.} \cite{FP4} explained how the geometrically understood Foucault pendulum can serve as a prototype for more advanced phenomena in physics known as Berry's phase or geometric phases. Capistrano \emph{et al.} \cite{PP1} investigated the anomalous planets precession in the nearly-Newtonian gravitational regime. Xu \emph{et al.} \cite{PP2} derived the analytic solution of a planetary orbit disturbed by the solar gravitational oblateness. Barker \emph{et al.} \cite{PP3} studied the turbulent damping of precessional fluid motions, in the simplest local computational model of a giant planet (or star), with and without a weak internal magnetic field. Nyambuya \cite{PP4} presented an improved version of the azimuthally symmetric theory of gravitation model, which is able to explain the anomalous perihelion precession. Sultana \emph{et al.} \cite{PP5} investigated the perihelion shift of planetary motion in conformal Weyl gravity using the metric of the static and spherically symmetric solution. Zhao \emph{et al.} \cite{PP6} found the exact solution to geodesic equation in the braneworld black hole spacetime by means of Jacobian elliptic function, obtaining the angle of perihelion precession from the zero point of Jacobian elliptic function for Mercury. Xu \cite{PP7} analyzed the perihelion precessions of orbit with arbitrary eccentricity, making use of the Laplace-Runge-Lenz vector. Zhang \emph{et al.} \cite{PP8} ascribed the twin kilohertz Quasi Periodic Oscillations of X-ray spectra to the pseudo-Newtonian Keplerian frequency and the apogee and perigee precession frequency. Saadat \emph{et al.} \cite{PP9} visualized the effect of dark matter on solar system and especially perihelion precession of Earth planet.

The field-induced precession, especially the Larmor precession, can be applied to investigate the physical properties of the nano-particles, crystal lattice, and neutron beam and so forth. Bagryansky \emph{et al.} \cite{LP1} studied the singlet-triplet oscillations in spin-correlated radical pairs, at magnetic field strengths low for one radical and high for the other. Home \emph{et al.} \cite{LP2} provided a detailed analysis for solving the appropriate Schr\"{o}dinger equation, for a spin-polarized plane wave passing through a spin-rotator containing uniform magnetic field. Janosfalvi \emph{et al.} \cite{LP3} described the numerous phenomenological equations used in the study of the behavior of single-domain magnetic nano-particles. Li \emph{et al.} \cite{LP4} investigated the Larmor precession of a neutral spinning particle in a magnetic field confined to the region of a one-dimensional rectangular barrier. Guo \emph{et al.} \cite{LP5} proposed an experimental method to detect the Larmor precession of a single spin with a spin-polarized tunneling current. Muller \emph{et al.} \cite{LP6} probed the dynamics of a single hole spin in a single, electrically tunable self-assembled quantum dot molecule formed by vertically stacking (In,Ga)As quantum dots. Rekveldt \emph{et al.} \cite{LP7} presented a method of carrying out high-precision measurements of crystal lattice parameters with Larmor precession. Mizukami \emph{et al.} \cite{LP8} investigated the magnetization precessions in the epitaxial films grown on MgO substrates, by means of an all-optical pump-probe method. Rekveldt \emph{et al.} \cite{LP9} considered the magnetized foils as the flippers to vary the neutron wavelength. The authors \cite{LP10} reviewed the techniques situation in which the Larmor precession has been used in the neutron spin-echo and neutron depolarization. Hautmann \cite{LP11} applied the circularly polarized light to inject partially spin-polarized electrons and holes in bulk germanium via both direct and indirect optical transitions. Bouwman \cite{LP12} modulated the intensity of a neutron beam using Larmor precession techniques.

After a careful and contrastive analysis of the above studies, one is able to find several major problems relevant to the precessional angular velocity as follows.

a) Physical quantity. The force may generate the revolution in the curvilinear motion, while the torque is capable of producing the rotation in the curvilinear motion. Similarly, besides the torque, the external magnetic flux density is able to induce the precessional angular velocity as well. There may be many sorts of physical quantities capable of leading to the precessional angular velocities.

b) Precessional motion. The force is different from the torque. So the type of curvilinear motion caused by the force can be considered to be independent of that caused by the torque. Analogously, the sort of precessional angular velocity induced by the external magnetic flux density will be considered to be different from that induced by the torque. Is it possible that there are some different types of precessional motions?

c) Orientation of precession. The existing classical mechanics and quantum mechanics both are unable to account for a few conundrums relevant to the orientation of precession. For instance, why is the magnetic moment of a boson different from that of a fermion, in the magnetic fields? why does the charged particle often revolve around the external magnetic flux density?

Forming a striking contrast to the above is that it is capable of resolving a few problems, derived from the classical mechanics and even quantum mechanics, in the complex-octonion electromagnetic and gravitational theories \cite{weng1} , attempting to improve some theoretical explanations, associated with the precessional angular velocity and the orientation of precession, to a certain extent.

In 1843, W. R. Hamilton invented the algebra of quaternions. Later the algebra of octonions was discovered independently by J. T. Graves and A. Cayley. The quaternion and octonion with the real coordinate values are the classical quaternion and octonion respectively. Further, in case a part of coordinate values of the quaternion and octonion are imaginary numbers and even complex numbers, they are called as the complex-quaternion and complex-octonion respectively.

J. C. Maxwell utilized firstly the real quaternions to expatiate the electromagnetic theory. His research methodology inspired subsequent scholars to employ the real and even complex octonions to explore the electromagnetic fields \cite{tanisli1, tanisli2}, gravitational fields \cite{demir1}, Dirac wave equation \cite{deleo}, general relativity, dark matter \cite{chanyal1}, multi-fluid plasma \cite{demir2, demir3}, dyons, and astrophysical jet \cite{weng2} and so on.

The algebra of quaternions is able to express effectively the precessional angular velocity. In the rotation of a rigid body with respect to one fixed point, the vector terminology can be applied to describe the Eulerian angles, including the nutation angle, precession angle, and intrinsic rotation angle. Making use of the vector terminology to depict the Eulerian angles of the satellite orbits, a few crucial matrices may go wrong occasionally, that is, each of matrix elements of one crucial matrix is equal to zero simultaneously. But this kind of failure will never happen again, when one utilizes the quaternions to express the Eulerian angles of the satellite orbits. By comparison with the vector terminology, the quaternions have obvious advantages to describe the Eulerian angles.

According to the precession equilibrium equation, the external field strength is able to induce the precessional angular velocity of particles, inferring the angular velocity of Larmor precession and some new predictions. However, the precessional angular velocity, induced by the external field strengths, is distinct from that generated by the gyroscopic torque in the theoretical mechanics. Especially, even if the orbital angular momentum is zero, the external electromagnetic strength is still able to induce a new term of precessional angular velocity and a new term of precessional angular momentum, producing a new ingredient of orbital angular momentum. It states that the precessional angular momentum would just be one special component of orbital angular momentum, in terms of the complex-octonion spaces \cite{weng3}.

In the octonionic electromagnetic and gravitational theories, the precessional angular velocity possesses several significant characteristics as follows.

a) Torque derivative. The torque is able to generate the precessional angular velocity, while the torque derivative (the vector, rather than the force) is capable of inducing the precessional angular velocity as well. Some ingredients of the torque derivative may be in direct proportion to the external field strengths. Apparently, the precessional angular velocity, induced by the torque derivative, is distinct from that generated by the torque. They are two different types of precessional angular velocities.

b) Precessional axis. When one component of the torque derivative is capable of playing a major role in the precessional motion, the precessional axis of the particle will orient this constituent of the torque derivative. Especially, when the contribution of external field strength plays a major role in the precessional motion, the particle will revolve around the external field strength. In general, the orientation of precession should be composite and intricate.

c) Spatial dimension. The precessional angular velocity, induced by the torque derivative, is related with the curl of linear velocity of the precessional motion. Further the curl is relevant to the spatial dimension of the linear velocity. As a result, the precessional angular velocity is multiple-value, in the complex-quaternion spaces. Its value depends on the actual spatial dimension of the precessional motion.

In the paper, by means of the octonionic field potential and quaternion operator (in Section 2), one can define the octonionic field strength, field source, and linear momentum, in the electromagnetic or gravitational fields. Further we can infer the angular momentum, torque, and force. If the octonion force equals to zero, we can infer the precession equilibrium equation and force equilibrium equation and so forth. The angular velocity of precession for particles will be deduced from the precession equilibrium equation, in the external electromagnetic or gravitational fields. Meanwhile, the angular velocity of revolution for particles can be inferred from the force equilibrium equation. The research reveals that the precessional angular velocity, induced by the torque derivative, is multiple-value, and relevant to the spatial dimension of the precessional motion.

The octonions, introduced by J. T. Graves and A. Cayley independently, can be called as the standard octonions. What the paper discusses is the standard octonion, rather than the non-standard octonions, including split-octonions \cite{chanyal2, demir4}, pseudo-octonions, Cartan's octonions \cite{furui, baez}, and others. The standard octonions are capable of exploring the gravitational theory, electromagnetic theory, general relativity \cite{weng4, weng5}, and quantum mechanics and so forth. The different types of octonions (standard octonions and non-standard octonions) are able to describe different kinds of physical properties. In the mathematics, each non-standard octonion can be considered as the function of the standard octonions. It means that the contribution from the non-standard octonions may not conflict with that of the standard octonions. The standard octonions and non-standard octonions \cite{mironov} can be combined together to become a whole, to depict jointly various physical properties, including the force and precessional angular velocity and so forth.

When the octonion force equals to zero, it is able to generate eight independent equations. Only three of them have explicit physical meanings, including the current continuity equation, fluid continuity equation, and force equilibrium equation. Nowadays, we found the clear physical meaning of the fourth independent equation, which associated with the physical property of precessional angular velocities for several particles. That is what we study in the paper. And the fourth independent equation is called as the `precession equilibrium equation' temporarily.

\section{Octonion force}

According to the basic postulates and the algebra of octonions in Ref.\cite{weng4}, we can define the octonionic field potential, field strength, field source, and linear momentum. Furthermore, one may deduce the octonionic angular momentum, torque, and force and so on as well.

\subsection{Angular momentum}

In the gravitational fields, the space is chosen as the complex-quaternion space $\mathbb{H}_g$, in which the basis vector is $\textbf{\emph{i}}_j$, the radius vector is $\mathbb{R}_g = i r_0 \textbf{\emph{i}}_0 + \Sigma r_k \textbf{\emph{i}}_k$ , and the velocity is $\mathbb{V}_g = i v_0 \textbf{\emph{i}}_0 + \Sigma v_k \textbf{\emph{i}}_k$. The gravitational potential is $\mathbb{A}_g = i a_0 \textbf{\emph{i}}_0 + \Sigma a_k \textbf{\emph{i}}_k$, the gravitational strength is $\mathbb{F}_g = f_0 \textbf{\emph{i}}_0 + \Sigma f_k \textbf{\emph{i}}_k$, and the gravitational source is $\mathbb{S}_g = i s_0 \textbf{\emph{i}}_0 + \Sigma s_k \textbf{\emph{i}}_k$ . Herein $\textbf{r} = \Sigma r_k \textbf{\emph{i}}_k$. $\textbf{v} = \Sigma v_k \textbf{\emph{i}}_k$. $\textbf{a} = \Sigma a_k \textbf{\emph{i}}_k$. $\textbf{f} = \Sigma f_k \textbf{\emph{i}}_k$. $\textbf{s} = \Sigma s_k \textbf{\emph{i}}_k$. The quaternionic operator is $\lozenge = i \textbf{\emph{i}}_0 \partial_0 + \Sigma \textbf{\emph{i}}_k \partial_k$. $\nabla = \Sigma \textbf{\emph{i}}_k \partial_k$, $\partial_j = \partial / \partial r_j$. $r_0 = v_0 t$. $v_0$ is the speed of light, and $t$ is the time. $r_j$ , $v_j$ , $a_j$, $s_j$, and $f_0$ are all real. $i$ is the imaginary unit. $f_k$ is the complex-number. $\textbf{\emph{i}}_k^2 = - 1$. $\textbf{\emph{i}}_0 = 1$. $k = 1, 2, 3$. $j = 0, 1, 2, 3$.

In the electromagnetic fields, the space is chosen as the complex 2-quaternion (short for the second quaternion) space $\mathbb{H}_e$ , in which the basis vector is $\textbf{\emph{I}}_j$ , the radius vector is $\mathbb{R}_e = i R_0 \textbf{\emph{I}}_0 + \Sigma R_k \textbf{\emph{I}}_k$, and the velocity is $\mathbb{V}_e = i V_0 \textbf{\emph{I}}_0 + \Sigma V_k \textbf{\emph{I}}_k$ . The electromagnetic potential is $\mathbb{A}_e = i A_0 \textbf{\emph{I}}_0 + \Sigma A_k \textbf{\emph{I}}_k$, the electromagnetic strength is $\mathbb{F}_e = F_0 \textbf{\emph{I}}_0 + \Sigma F_k \textbf{\emph{I}}_k$, and the electromagnetic source is $\mathbb{S}_e = i S_0 \textbf{\emph{I}}_0 + \Sigma S_k \textbf{\emph{I}}_k$ . Herein $\textbf{R}_0 = R_0 \textbf{\emph{I}}_0$ . $\textbf{V}_0 = V_0 \textbf{\emph{I}}_0$. $\textbf{A}_0 = A_0 \textbf{\emph{I}}_0$. $\textbf{F}_0 = F_0 \textbf{\emph{I}}_0$ . $\textbf{S}_0 = S_0 \textbf{\emph{I}}_0$. $\textbf{R} = \Sigma R_k \textbf{\emph{I}}_k$. $\textbf{V} = \Sigma V_k \textbf{\emph{I}}_k$. $\textbf{A} = \Sigma A_k \textbf{\emph{I}}_k$. $\textbf{F} = \Sigma F_k \textbf{\emph{I}}_k$. $\textbf{S} = \Sigma S_k \textbf{\emph{I}}_k$ . $\mathbb{H}_e = \mathbb{H}_g \circ \textbf{\emph{I}}_0$. $\textbf{\emph{I}}_j^2 = - 1$. $\circ$ denotes the multiplication of octonions. $F_k$ is the complex-number. $R_j$ , $V_j$ , $A_j$ , $S_j$, and $F_0$ are all real.

As a result, two independent spaces, $\mathbb{H}_g$ and $\mathbb{H}_e$ , can be considered to be perpendicular to each other. It means that both of them can be combined together to become one single complex-octonion space $\mathbb{O}$ , which is fitting for describing the physical properties of the electromagnetic and gravitational fields. In the complex-octonion space $\mathbb{O}$ , the octonionic field source $\mathbb{S}$ can be defined as,
\begin{eqnarray}
\mu \mathbb{S} && = - ( i \mathbb{F} / v_0 + \lozenge )^\ast \circ \mathbb{F}
\nonumber
\\
&& = \mu_g \mathbb{S}_g + k_{eg} \mu_e \mathbb{S}_e - i \mathbb{F}^\ast \circ \mathbb{F} / v_0  ~ ,
\end{eqnarray}
where $\mathbb{S}_e = q \mathbb{V}_e$ , and $\mathbb{S}_g = m \mathbb{V}_g$ , for one particle. The symbol $\ast$ stands for the octonion conjugate. $q$ is the density of electric charge, while $m$ is the density of inertial mass. $\mu_g < 0$, and $\mu_e > 0$. $k_{eg}$ , $\mu$ , $\mu_g$ , and $\mu_e$ are coefficients.

From the above, the octonionic linear momentum can be defined as,
\begin{eqnarray}
\mathbb{P} = \mathbb{P}_g + k_{eg} \mathbb{P}_e   ~ ,
\end{eqnarray}
where $\mathbb{P}_e = \mu_e \mathbb{S}_e / \mu_g$ . $\mathbb{P}_e = i \textbf{P}_0 + \textbf{P}$ , $\textbf{P} = \Sigma P_k \textbf{\emph{I}}_k$ , $\textbf{P}_0 = P_0 \textbf{\emph{I}}_0$ . Meanwhile $\mathbb{P}_g = \{\mu_g \mathbb{S}_g - ( i \mathbb{F} / v_0 )^\ast \circ \mathbb{F} \} / \mu_g$ . $\mathbb{P}_g = i p_0 + \textbf{p}$ , $\textbf{p} = \Sigma p_k \textbf{\emph{i}}_k$ . $p_j$ and $P_j$ are all real.

The octonionic angular momentum $\mathbb{L}$ , in the complex-octonion spaces, can be derived from the linear momentum $\mathbb{P}$ and the radius vector, $\mathbb{R} = \mathbb{R}_g + k_{eg} \mathbb{R}_e$, and so forth (see Ref.\cite{weng1}). Furthermore the octonion angular momentum $\mathbb{L}$ , in the electromagnetic and gravitational fields, can be separated into, $\mathbb{L} = \mathbb{L}_g + k_{eg} \mathbb{L}_e$ . Herein $\mathbb{L}_g = L_{10} + i \textbf{L}_1^i + \textbf{L}_1$. $\textbf{L}_1 = \Sigma L_{1k} \textbf{\emph{i}}_k$, $\textbf{L}_1^i = \Sigma L_{1k}^i \textbf{\emph{i}}_k$ . $\mathbb{L}_e = \textbf{L}_{20} + i \textbf{L}_2^i + \textbf{L}_2$. $\textbf{L}_{20} = L_{20} \textbf{\emph{I}}_0$ . $\textbf{L}_2 = \Sigma L_{2k} \textbf{\emph{I}}_k$ , $\textbf{L}_2^i = \Sigma L_{2k}^i \textbf{\emph{I}}_k$. $L_{1j}$ , $L_{1k}^i$ , $L_{2j}$ , and $L_{2k}^i$ are all real.

\subsection{Force}

From the above, it is able to define the octonionic torque $\mathbb{W}$ as follows,
\begin{eqnarray}
\mathbb{W} = - v_0 ( i \mathbb{F} / v_0 + \lozenge ) \circ \mathbb{L}   ~ , \label{equ:3}
\end{eqnarray}
where $\mathbb{W} = \mathbb{W}_g + k_{eg} \mathbb{W}_e$ . $\mathbb{W}_g = i W_{10}^i + W_{10} + i \textbf{W}_1^i + \textbf{W}_1$ . $\mathbb{W}_e = i \textbf{W}_{20}^i + \textbf{W}_{20} + i \textbf{W}_2^i + \textbf{W}_2$. $W_{10}^i$ is the energy. $W_{10}$ is the divergence of angular momentum. $- \textbf{W}_1^i$ is the torque. $\textbf{W}_1$ is the curl of angular momentum. $\textbf{W}_{20}^i$ is the second-energy, containing the divergence of electric moment. $\textbf{W}_2^i$ is the second-torque, comprising the curl of electric moment and the derivative of magnetic moment. $\textbf{W}_{20}$ includes the divergence of magnetic moment. $\textbf{W}_2$ covers the curl of magnetic moment and the derivative of electric moment. $\textbf{W}_1 = \Sigma W_{1k} \textbf{\emph{i}}_k$, $\textbf{W}_1^i = \Sigma W_{1k}^i \textbf{\emph{i}}_k$ . $\textbf{W}_{20} = W_{20} \textbf{\emph{I}}_0$, $\textbf{W}_{20}^i = W_{20}^i \textbf{\emph{I}}_0$ . $\textbf{W}_2 = \Sigma W_{2k} \textbf{\emph{I}}_k$ , $\textbf{W}_2^i = \Sigma W_{2k}^i \textbf{\emph{I}}_k$ . $W_{1j}$ , $W_{1j}^i$ , $W_{2j}$ , and $W_{2j}^i$ are all real.

From the above, the octonionic force $\mathbb{N}$ can be defined as,
\begin{eqnarray}
\mathbb{N} = - ( i \mathbb{F} / v_0 + \lozenge ) \circ \mathbb{W}   ~ ,
\label{equ:4}
\end{eqnarray}
where $\mathbb{N} = \mathbb{N}_g + k_{eg} \mathbb{N}_e$ . $\mathbb{N}_g = i N_{10}^i + N_{10} + i \textbf{N}_1^i + \textbf{N}_1$ . $\mathbb{N}_e = i \textbf{N}_{20}^i + \textbf{N}_{20} + i \textbf{N}_2^i + \textbf{N}_2$. $N_{10}$ is the power. $N_{10}^i$ comprises the divergence of torque. $\textbf{N}_1$ is the torque derivative. $\textbf{N}_1^i$ is the force. $\textbf{N}_{20}$ is the second-power, which is similar to $N_{10}$ . $\textbf{N}_{20}^i$ contains the term $(\partial_0 \textbf{W}_{20} + \nabla \cdot \textbf{W}_2^i )$. $\textbf{N}_2$ includes the $(- \nabla \circ \textbf{W}_{20} - \nabla \times \textbf{W}_2 + \partial_0 \textbf{W}_2^i )$. $\textbf{N}_2^i$ is the second-force, embodying the term $(\partial_0 \textbf{W}_2 + \nabla \circ \textbf{W}_{20}^i + \nabla \times \textbf{W}_2^i )$. $\textbf{N}_1 = \Sigma N_{1k} \textbf{\emph{i}}_k$, $\textbf{N}_1^i = \Sigma N_{1k}^i \textbf{\emph{i}}_k$. $\textbf{N}_2 = \Sigma N_{2k} \textbf{\emph{I}}_k$, $\textbf{N}_{20} = N_{20} \textbf{\emph{I}}_0$ . $\textbf{N}_2^i = \Sigma N_{2k}^i \textbf{\emph{I}}_k$ , $\textbf{N}_{20}^i = N_{20}^i \textbf{\emph{I}}_0$ . $N_{1j}$ , $N_{1j}^i$ , $N_{2j}$, and $N_{2j}^i$ are all real.

\begin{table}[h]
\centering
\caption{In the complex-octonion spaces, there are two different types of precessional angular velocities, caused by two distinct physical quantities respectively. One obeys the formula of octonion torque, while the other meets the demand of the equation of octonion force.}
\label{tab:1}       % Give a unique label
\begin{tabular}{lll}
\hline\noalign{\smallskip}
precessional angular velocity   &   formula                                                                   &   special case                                     \\
\noalign{\smallskip}\hline\noalign{\smallskip}
torque-induced                  &   $\mathbb{W} = - v_0 ( i \mathbb{F} / v_0 + \lozenge ) \circ \mathbb{L} $  &   $\textbf{W}_1^i = - v_0 \partial_0 \textbf{L}$   \\
field-induced                   &   $\mathbb{N} = - ( i \mathbb{F} / v_0 + \lozenge ) \circ \mathbb{W}$       &   $\textbf{N}_1 = 0$                               \\
\noalign{\smallskip}\hline
\end{tabular}
\end{table}

\section{Equilibrium equations}

In the complex-octonion spaces, there exists the octonion force equation, $\mathbb{N} = 0$, under certain circumstances. This equation can be separated into eight independent equilibrium/continuity equations in the electromagnetic and gravitational fields. Four of them possess explicit physical meanings, including the precession equilibrium equation, current continuity equation, force equilibrium equation, and fluid continuity equation (Table \ref{tab:2}). The continuity equations and equilibrium equations are the same essentially, in the complex-octonion space $\mathbb{O}$ .

\subsection{Force equilibrium equation}

In the complex-quaternion space $\mathbb{H}_g$ , from the expansion of Eq.(\ref{equ:4}), the definition of force, $\textbf{N}_1^i$ , can be written as,
\begin{eqnarray}
\textbf{N}_1^i = && ( W_{10}^i \textbf{g} / v_0 + \textbf{g} \times \textbf{W}_1^i / v_0 - W_{10} \textbf{b} - \textbf{b} \times \textbf{W}_1 ) / v_0
\nonumber
\\
&&
+ k_{eg}^2 ( \textbf{E} \circ \textbf{W}_{20}^i / v_0 + \textbf{E} \times \textbf{W}_2^i / v_0  ) / v_0
\nonumber
\\
&&
- k_{eg}^2 ( \textbf{B} \circ \textbf{W}_{20} + \textbf{B} \times \textbf{W}_2 ) / v_0
\nonumber
\\
&&
- (\partial_0 \textbf{W}_1 + \nabla W_{10}^i + \nabla \times \textbf{W}_1^i )  ~ ,
\end{eqnarray}
where $\textbf{B}$ is the magnetic flux density, while $\textbf{E}$ is the electric field intensity. $\textbf{g}$ is the gravitational acceleration. And $\textbf{b}$ is the gravitational precessional-angular-velocity, for it is relevant to the precessional angular velocity (see Ref.\cite{weng3}).

When the field strength $\mathbb{F}$ is comparatively weak, the above will approximate to,
\begin{eqnarray}
\textbf{N}_1^i / k_p \approx   &&   p_0 \textbf{g} / v_0  + k_{eg}^2 ( \textbf{E} \circ \textbf{P}_0 / v_0 - \textbf{B} \times \textbf{P} ) - \textbf{b} \times \textbf{p} - \nabla (p_0 v_0)
\nonumber
\\
&&
+ L_{10} ( \textbf{g} \times \textbf{b} + k_{eg}^2 \textbf{E} \times \textbf{B} ) / ( v_0^2 k_p )
- \partial_0 (\textbf{p} v_0)   ~ ,
\end{eqnarray}
where $k_{eg}^2 ( \textbf{E} \circ \textbf{P}_0 / v_0 - \textbf{B} \times \textbf{P} )$ is the electromagnetic force. $( p_0 \textbf{g} / v_0 )$ is the gravity. $\partial_0 ( - \textbf{p} v_0)$ is the inertial force. $\nabla ( - p_0 v_0)$ is the energy gradient. ${ L_{10} (k_{eg}^2 \textbf{E} \times \textbf{B} ) / ( v_0^2 k_p ) }$ is in direct proportion to the electromagnetic momentum. $p_0 = m^\prime v_0$ , $\textbf{p} = m \textbf{v}$ . $m^\prime$ is the density of gravitational mass. $k_p = (k - 1)$ is one coefficient, with $k$ being the spatial dimension of the vector $\textbf{r}$ .

When the force, $\textbf{N}_1^i$ , equals to zero, it is able to deduce the conventional force equilibrium equation from the above,
\begin{eqnarray}
\textbf{N}_1^i = 0   ~ ,
\end{eqnarray}
and it is revealed that $k_{eg}^2 = \mu_g / \mu_e < 0$.

\subsection{Fluid continuity equation}

In the complex-quaternion space $\mathbb{H}_g$ , from the expansion of Eq.(\ref{equ:4}), the power, $N_{10}$, can be defined as,
\begin{eqnarray}
N_{10} = && ( \textbf{g} \cdot \textbf{W}_1 / v_0 + \textbf{b} \cdot \textbf{W}_1^i ) / v_0
+ (\partial_0 W_{10}^i - \nabla \cdot \textbf{W}_1 )
\nonumber
\\
&&
+ k_{eg}^2 ( \textbf{E} \cdot \textbf{W}_2 / v_0 + \textbf{B} \cdot \textbf{W}_2^i ) / v_0
~ ,
\end{eqnarray}
further, when the field strength $\mathbb{F}$ is comparatively weak, the above will be reduced into,
\begin{eqnarray}
N_{10} / k_p  \approx  && \partial_0 ( p_0 v_0 ) - \nabla \cdot ( \textbf{p} v_0 )
+ ( \textbf{g} \cdot \textbf{p} + k_{eg}^2 \textbf{E} \cdot \textbf{P} ) / v_0
\nonumber
\\
&&
+ L_{10} ( \textbf{b} \cdot \textbf{b} - \textbf{g} \cdot \textbf{g} / v_0^2 ) / ( v_0 k_p )
\nonumber
\\
&&
+ k_{eg}^2 L_{10} ( \textbf{B} \cdot \textbf{B} - \textbf{E} \cdot \textbf{E} / v_0^2 ) / ( v_0 k_p )
~ ,
\end{eqnarray}
where $\textbf{W}_2^i \approx \textbf{v} \times \textbf{P}$ . $\textbf{W}_1 \approx k_p \textbf{p} v_0$ . $\textbf{W}_2 \approx k_p \textbf{P} v_0$ .

When the power equals to zero, one can achieve the fluid continuity equation as follows,
\begin{eqnarray}
N_{10} = 0  ~ ,
\end{eqnarray}
where the power covers the term, $( \textbf{g} \cdot \textbf{p} + k_{eg}^2 \textbf{E} \cdot \textbf{P} )$ , which is able to translate into the Joule heat. And it has an influence on the fluid continuity equation directly.

In case there is no field strength, the above will be reduced into the conventional fluid continuity equation,
\begin{eqnarray}
\partial_0 p_0 - \nabla \cdot \textbf{p} = 0  ~ .
\end{eqnarray}

Essentially, the fluid continuity equation may merely be one type of equilibrium equation, from another point of view.

\subsection{Current continuity equation}

From the expansion of Eq.(\ref{equ:4}), in the complex-quaternion space $\mathbb{H}_g$ , the second-power, $\textbf{N}_{20}$, can be defined as,
\begin{eqnarray}
\textbf{N}_{20} = && ( \textbf{g} \cdot \textbf{W}_2 / v_0 + \textbf{b} \cdot \textbf{W}_2^i ) / v_0
+ (\partial_0 \textbf{W}_{20}^i - \nabla \cdot \textbf{W}_2 )
\nonumber
\\
&&
+ ( \textbf{E} \cdot \textbf{W}_1 / v_0 + \textbf{B} \cdot \textbf{W}_1^i ) / v_0     ~ ,
\end{eqnarray}
further, when the field strength $\mathbb{F}$ is comparatively weak, the above will be simplified into,
\begin{eqnarray}
\textbf{N}_{20} / k_p   \approx  && \partial_0 ( \textbf{P}_0 v_0 ) - \nabla \cdot ( \textbf{P} v_0 )
+ \textbf{g} \cdot \textbf{P} / v_0 + \textbf{E} \cdot \textbf{p} / v_0
\nonumber
\\
&&
+ ( \textbf{b} \cdot \textbf{b} + \textbf{B} \cdot \textbf{B} ) \textbf{L}_{20} / ( v_0 k_p )
~ ,
\end{eqnarray}
where each component of the field strength $\mathbb{F}$ makes a contribution to the above.

When the second-power, $\textbf{N}_{20}$ , equals to zero, it is able to deduce the current continuity equation as,
\begin{eqnarray}
\textbf{N}_{20} = 0     ~ ,
\end{eqnarray}
where the above includes the cross-term, $( \textbf{g} \cdot \textbf{P} + \textbf{E} \cdot \textbf{p} ) / v_0 $ , between the electromagnetic field and gravitational field. And the field strength $\mathbb{F}$ and linear momentum $\mathbb{P}$ and so forth will impact the current continuity equation.

When there is no field strength, the above will be degenerated into the conventional current continuity equation ,
\begin{eqnarray}
\partial_0  \textbf{P}_0  - \nabla \cdot \textbf{P}  = 0     ~ ,
\end{eqnarray}
and the current continuity equation is essentially one type of equilibrium equation, from the viewpoint of the basis vector $\textbf{I}_0$ , in the complex-octonion space $\mathbb{O}$ .

Besides the above three equilibrium equations, there exist other types of equilibrium equations also, especially the precession equilibrium equation in the next section. From the single octonion equation, $\mathbb{N} = 0$ , it is able to deduce simultaneously the fluid continuity equation, force equilibrium equation, current continuity equation, and precession equilibrium equation. In case the first three equations are all tenable, the last one must be tenable as well. It means that the precession equilibrium equation has a solid theoretical foundation, in the complex-octonion spaces.

\begin{table}[h]
\centering
\caption{The octonion force equation, $\mathbb{N} = 0$, can be separated into eight equilibrium/continuity equations independent of each other. Especially the four equilibrium/continuity equations possess explicit physical meanings, in the paper. Apparently, the continuity equations belong to the equilibrium equations essentially, in the complex-octonion spaces.}
\label{tab:2}       % Give a unique label
\begin{tabular}{lll}
\hline\noalign{\smallskip}
equilibrium/continuity equation                    &   definition                              &  space           \\
\noalign{\smallskip}\hline\noalign{\smallskip}
force equilibrium equation                         &   $\textbf{N}_1^i = 0$                    &  $\mathbb{H}_g$  \\
fluid continuity equation                          &   $N_{10} = 0$                            &  $\mathbb{H}_g$  \\
precession equilibrium equation                    &   $\textbf{N}_1 = 0$                      &  $\mathbb{H}_g$  \\
torque continuity equation                         &   $N_{10}^i = 0$                          &  $\mathbb{H}_g$  \\
current continuity equation                        &   $\textbf{N}_{20} = 0$                   &  $\mathbb{H}_e$  \\
second-precession equilibrium equation             &   $\textbf{N}_2 = 0$                      &  $\mathbb{H}_e$  \\
second-force equilibrium equation                  &   $\textbf{N}_2^i = 0$                    &  $\mathbb{H}_e$  \\
second-torque continuity equation                  &   $\textbf{N}_{20}^i = 0$                 &  $\mathbb{H}_e$  \\
\noalign{\smallskip}\hline
\end{tabular}
\end{table}

\section{Precession equilibrium equation}

When the octonion force $\mathbb{N}$ equals to zero, in the complex-octonion space $\mathbb{O}$ , it is capable of inferring eight independent equilibrium/continuity equations. One of them is the precession equilibrium equation, $\textbf{N}_1 = 0$ . Further, it is able to deduce the angular velocities of precession for charged or neutral particles, from the precession equilibrium equation, $\textbf{N}_1 = 0$ . This equilibrium equation will be disturbed by the electromagnetic strength, gravitational strength, spatial dimension, and torque and so forth.

The precession equilibrium equation, $\textbf{N}_1 = 0$ , can be expanded into,
\begin{eqnarray}
0 = && ( W_{10} \textbf{g} / v_0 + \textbf{g} \times \textbf{W}_1 / v_0 + W_{10}^i \textbf{b} + \textbf{b} \times \textbf{W}_1^i ) / v_0
\nonumber
\\
&&
+ k_{eg}^2 ( \textbf{E} \circ \textbf{W}_{20} / v_0 + \textbf{E} \times \textbf{W}_2 / v_0  ) / v_0
\nonumber
\\
&&
+ k_{eg}^2 (  \textbf{B} \circ \textbf{W}_{20}^i + \textbf{B} \times \textbf{W}_2^i ) / v_0
\nonumber
\\
&&
+ ( \partial_0 \textbf{W}_1^i - \nabla W_{10} - \nabla \times \textbf{W}_1 )~ ,
\label{equ:16}
\end{eqnarray}
where the term in the above is the torque-derivative term, rather than the force term, although both of which are the vectors and possess the same dimension. Obviously the field strength and others may exert an influence on the precession equilibrium equation (Table \ref{tab:3}).

In the complex-octonion spaces, the above can be reduced into a few special cases as follows.

\subsection{Torque-derivative term}

In case two terms, $\partial_0 \textbf{W}_1^i$ and $\nabla \times \textbf{W}_1$ , play a major role in the precessional motion, meanwhile there is no field strength, and other tiny terms can be neglected, the precession equilibrium equation, $\textbf{N}_1 = 0$, can be degenerated into,
\begin{eqnarray}
\partial_0 \textbf{W}_1^i - \nabla \times \textbf{W}_1 =  0     ~ ,
\label{equ:17}
\end{eqnarray}
where the torque-derivative term, $\partial_0 \textbf{W}_1^i$ , is relevant to some influence factors, according to the definition of torque. $\nabla \times \textbf{W}_1 \approx k_p m v_0 \nabla \times \textbf{v}$ . The linear velocity for the objects consists of the linear velocity caused by the motions of revolution, rotation, and precession. In the paper, we discuss merely the linear velocity $\textbf{v}_p$ caused by the precessional motion. As a result, $\nabla \times \textbf{v}_p = k \overrightarrow{\omega}_p$. $\overrightarrow{\omega}_p$ is the angular velocity of precession, while $k$ is the spatial dimension of linear velocity $\textbf{v}_p$ .

Further the above will be simplified into
\begin{eqnarray}
\partial_0 \textbf{W}_1^i - k_p m k v_0 \overrightarrow{\omega}_p =  0     ~ ,
\label{equ:18}
\end{eqnarray}
where the angular velocity of precession is related with the spatial dimension, mass, and torque-derivative term and so forth.

The above states that the torque-derivative term, $\partial_0 \textbf{W}_1^i$ , will produce the angular velocity of precession, even if there is no field strength.

\subsection{Magnetic field}

If two terms, $k_{eg}^2 \textbf{B} \circ \textbf{W}_{20}^i / v_0$ and $\nabla \times \textbf{W}_1$ , play a major role in the precessional motion, meanwhile there is no electric field intensity nor the gravitational strength, and other tiny terms can be neglected, the precession equilibrium equation, $\textbf{N}_1 = 0$, can be simplified into,
\begin{eqnarray}
k_{eg}^2 \textbf{B} \circ \textbf{W}_{20}^i / v_0 - \nabla \times \textbf{W}_1  =  0  ~ ,
\end{eqnarray}
where the second-energy $\textbf{W}_{20}^i$ is similar to the energy $W_{10}^i$ . $\textbf{W}_{20}^i \approx k_p \textbf{P}_0 v_0$ . $k_{eg}^2 \textbf{P}_0 = q V_0 \textbf{\emph{I}}_0$. $V_0 / v_0 \approx 1$.

Further the above will be reduced into,
\begin{eqnarray}
q \textbf{B} \circ \textbf{\emph{I}}_0 - k m \overrightarrow{\omega}_p = 0    ~ ,
\label{equ:20}
\end{eqnarray}
where the angular velocity of precession is, $\overrightarrow{\omega}_{p(1)} = q \textbf{B} \circ \textbf{\emph{I}}_0 / m$ , when $k = 1$. The angular velocity of precession is, $\overrightarrow{\omega}_{p(2)} = q \textbf{B} \circ \textbf{\emph{I}}_0 / (2 m)$ , when $k = 2$. And the angular velocity of precession is, $\overrightarrow{\omega}_{p(3)} = q \textbf{B} \circ \textbf{\emph{I}}_0 / (3 m)$ , when $k = 3$. According to the multiplication of octonions, the term, $\textbf{B} \circ \textbf{\emph{I}}_0$, is the vector in the complex-quaternion space $\mathbb{H}_g$ .

The above means that the magnetic flux density $\textbf{B}$ will induce the angular velocity of precession, revolving around the direction, $\textbf{B} \circ \textbf{\emph{I}}_0$ , for the charged objects. The inference can be applied to explicate the angular velocity of Larmor precession, for the charged particle and the spin angular momentum in certain magnetic fields (see Ref.\cite{weng3}). This will be helpful to understand the spin angular momentum relevant to the Zeeman effect in the quantum mechanics.

\subsection{Electric field}

When two terms, $k_{eg}^2 ( \textbf{E} \circ \textbf{W}_{20} ) / v_0^2$ and $\nabla \times \textbf{W}_1$ , play a major role in the precessional motion, meanwhile there is no magnetic flux density nor the gravitational strength, and other tiny terms can be neglected, the precession equilibrium equation, $\textbf{N}_1 = 0$, can be degraded into,
\begin{eqnarray}
k_{eg}^2  \textbf{E} \circ \textbf{W}_{20} / v_0^2  - \nabla \times \textbf{W}_1  = 0  ~ ,
\label{equ:21}
\end{eqnarray}
where the term $\textbf{W}_{20}$ is associated with the field strength and so forth. The term, $\textbf{E} \circ \textbf{\emph{I}}_0$ , is the vector in the complex-quaternion space $\mathbb{H}_g$ , according to the multiplication of octonions.

The above reveals that the electric field intensity $\textbf{E}$ will generate the angular velocity of precession, with the orientation of precession, $\textbf{E} \circ \textbf{\emph{I}}_0$ , for the charged objects. The term $\textbf{W}_{20}$ is more intricate than the second-energy $\textbf{W}_{20}^i$ , consequently the angular velocity of precession in the electric field intensity $\textbf{E}$ will be much more complicated than that in the magnetic flux density $\textbf{B}$ . And it can be utilized to unpuzzle some precessional phenomena of charged particles relevant to the Stark effect in the electric fields.

\subsection{Gravitational precessional-angular-velocity}

When two terms, $W_{10}^i \textbf{b} / v_0$ and $\nabla \times \textbf{W}_1$ , play a major role in the precessional motion, meanwhile there is no electromagnetic strength nor the gravitational acceleration, and other tiny terms can be neglected, the precession equilibrium equation, $\textbf{N}_1 = 0$ , can be degenerated into,
\begin{eqnarray}
W_{10}^i \textbf{b} / v_0 - \nabla \times \textbf{W}_1 = 0   ~ ,
\label{equ:22}
\end{eqnarray}
where the energy $W_{10}^i \approx k_p p_0 v_0$ . $m^\prime / m \approx 1$.

The above can be further simplified into,
\begin{eqnarray}
\textbf{b} - k \overrightarrow{\omega}_p = 0    ~ ,
\label{equ:23}
\end{eqnarray}
where the angular velocity of precession is, $\overrightarrow{\omega}_{p(1)} = \textbf{b} $ , when $k = 1$. The angular velocity of precession is, $\overrightarrow{\omega}_{p(2)} = \textbf{b} / 2$ , when $k = 2$. And the angular velocity of precession is, $\overrightarrow{\omega}_{p(3)} = \textbf{b} / 3$ , when $k = 3$.

The above states that the gravitational precessional-angular-velocity $\textbf{b}$ will induce the angular velocity of precession, with the orientation of precession, $\textbf{b}$ , for the neutral objects. And it can be applied to account for the dynamic properties of the astrophysical jets (see Ref.\cite{weng2}), including the precession, rotation, and collimation and so forth.

\subsection{Gravitational acceleration}

In case two terms, $ W_{10} \textbf{g} / v_0^2$ and $\nabla \times \textbf{W}_1$ , play a major role in the precessional motion, meanwhile there is no electromagnetic strength nor the gravitational precessional-angular-velocity $\textbf{b}$ , and other tiny terms can be neglected, the precession equilibrium equation, $\textbf{N}_1 = 0$ , can be degraded into,
\begin{eqnarray}
W_{10} \textbf{g} / v_0^2 - \nabla \times \textbf{W}_1 = 0    ~ ,
\label{equ:24}
\end{eqnarray}
where the term $W_{10}$ is relevant to the angular momentum and field strength and so forth.

The above states that the gravitational acceleration will result in the angular velocity of precession, with the orientation of precession, $\textbf{g}$ , for the neutral objects. The term $W_{10}$ is more involved than the energy $W_{10}^i$ , so the angular velocity of precession in the gravitational acceleration $\textbf{g}$ will be much more complicated than that in the gravitational precessional-angular-velocity $\textbf{b}$ . And it can be applied to explain some precessional phenomena of neutral particles in the gravitational field $\textbf{g}$ .

In the complex-octonion spaces, the torque and force may generate different types of curvilinear motions, for instance, the rotation and revolution. Similarly, the torque and torque-derivative can induce different types of precessional angular velocities, such as, the gyroscopic precession and field-induced precession. In the section, it should be noted that the precessional angular velocity, induced by the torque-derivative in Eq.(\ref{equ:4}), is quite a contrast to that derived from the octonion torque in Eq.(\ref{equ:3}).

\begin{table}[h]
\centering
\caption{The field-induced precession is a primary representative of the precession induced by the torque derivative. And the precessional angular velocity, induced by the external field strength, is relevant to the spatial dimension of precessional motion and the direction of field strength and so forth, in the complex-quaternion space $\mathbb{H}_g$ .}
\label{tab:3}       % Give a unique label
\begin{tabular}{lll}
\hline\noalign{\smallskip}
field strength  &  precessional angular velocity                                                                       &  orientation                                \\
\noalign{\smallskip}\hline\noalign{\smallskip}
$\textbf{B}$    &  $\overrightarrow{\omega}_{p(k)} = q \textbf{B} \circ \textbf{\emph{I}}_0 / ( m k) $                 &  $\textbf{B} \circ \textbf{\emph{I}}_0$     \\
$\textbf{E}$    &  $\overrightarrow{\omega}_{p(k)} = k_{eg}^2 \textbf{E} \circ \textbf{W}_{20}/( k_p m k v_0^3 ) $     &  $\textbf{E} \circ \textbf{\emph{I}}_0$     \\
$\textbf{b}$    &  $\overrightarrow{\omega}_{p(k)} = \textbf{b} / k $                                                  &  $\textbf{b}$                               \\
$\textbf{g}$    &  $\overrightarrow{\omega}_{p(k)} = W_{10} \textbf{g} / ( k_p m k v_0^3 ) $                           &  $\textbf{g}$                               \\
\noalign{\smallskip}\hline
\end{tabular}
\end{table}

\section{Precessional angular velocity}

In the complex-octonion spaces, some factors of the torque derivative may exert an impact on the precessional angular velocity, including the torque, partial derivative of torque, gravitational strength, and electromagnetic strength and so on. The influences of these factors on the precessional motions may be different from each other. So it is necessary to contrast and analyze various properties of the precessional angular velocities, especially the orientation of precession, relevant equations, and spatial dimension and so forth.

\subsection{Orientation of precession}

From Eq.(\ref{equ:18}), it is found that the partial derivative of torque, with respect to time, is able to induce directly the precessional angular velocity. If the torque derivative term, $v_0 \partial_0 \textbf{W}_1^i$ , is periodic, the precessional angular velocity, $\overrightarrow{\omega}_p$ , will be periodic accordingly. And that the period of precessional angular velocity is identical with that of the torque derivative term. Meanwhile the direction of precessional angular velocity orients that of the torque derivative term.

For the electromagnetic fields, either of two components, $\textbf{B}$ and $\textbf{E}$ , is capable of inducing the precessional angular velocity directly. From Eq.(\ref{equ:20}), it is found that the precessional angular velocity is, $\overrightarrow{\omega}_{p(k)} = q \textbf{B} \circ \textbf{\emph{I}}_0 / (m k)$ , in the magnetic flux density $\textbf{B}$ . And the direction of precessional angular velocity orients that of the vector $\textbf{B} \circ \textbf{\emph{I}}_0$ . From Eq.(\ref{equ:21}), the precessional angular velocity is comparatively complicated, and relevant to various factors, in the electric field intensity $\textbf{E}$ . And the direction of precessional angular velocity is along that of the vector $\textbf{E} \circ \textbf{\emph{I}}_0$ .

In terms of the gravitational fields, either of two constituents, $\textbf{g}$ and $\textbf{b}$, is able to induce the precessional angular velocity directly. According to Eq.(\ref{equ:23}), it is found that the precessional angular velocity is, $\overrightarrow{\omega}_{p(k)} = \textbf{b} / k$ , in the gravitational precessional-angular-velocity $\textbf{b}$ . And the direction of precessional angular velocity orients that of the vector $\textbf{b}$ . From Eq.(\ref{equ:24}), the precessional angular velocity is also comparatively complicated, and relevant to assorted factors, in the gravitational acceleration $\textbf{g}$ . And the direction of precessional angular velocity is along that of the vector $\textbf{g}$ , in the complex-quaternion space $\mathbb{H}_g$ .

Besides these factors in the above, certain other factors in Eq.(\ref{equ:16}) can directly lead to the precessional angular velocities as well.

\subsection{Spatial dimension}

According to Eq.(\ref{equ:17}), the precessional angular velocity is relevant to the curl of linear velocity of the precessional motion, while the curl is relevant to the spatial dimension $k$ of the linear velocity. In the complex-quaternion space $\mathbb{H}_g$ , the possible value of spatial dimension is, $k = 1, 2, 3$. Consequently, the precessional angular velocity may possess three types of possible values.

In terms of the charged particles in the magnetic field $\textbf{B}$ , the precessional angular velocity occupies three types of possible values, from Eq.(\ref{equ:20}). When $k = 1$, the precessional angular velocity corresponds to the one-dimensional precessional motion. It is possible to speculate that the precessional motion of Boson particles is one-dimensional. If $k = 2$, the precessional angular velocity may correspond to the two-dimensional precessional motion. We can speculate that the precessional motion of Fermion particles is two-dimensional. In case $k = 3$, the precessional angular velocity must correspond to the three-dimensional precessional motion of some particles. It is interesting to speculate that the possible values of the precessional angular velocities should be altered, in case the Bosons or Fermions and other particles were compelled to experience the three-dimensional precessional motions, in the magnetic fields.

Similarly, the precessional angular velocity has three types of possible values, for the neutral particles in the gravitational field $\textbf{b}$ . In case the precessional motion of neutral particles is one-dimensional, the value of the precessional angular velocity corresponds to the case $k = 1$. When the precessional motion of neutral particles is two-dimensional, the value of the precessional angular velocity may correspond to the case $k = 2$. If the precessional motion of some neutral particles is three-dimensional, the value of the precessional angular velocity will correspond to the case $k = 3$. In other words, the spatial dimension of precessional motion exerts an impact on the possible values of precessional angular velocity, for the neutral particles in the gravitational field $\textbf{b}$ .

In the electric field $\textbf{E}$ or gravitational field $\textbf{g}$ , there exist three possible values of precessional angular velocities as well. Furthermore, from Eq.(\ref{equ:16}), the spatial dimension of precessional motion of particles will also make a contribution to the possible values of precessional angular velocities, caused by the other physical quantities, besides the above field strengths.

\subsection{Two equations}

In the curvilinear motions, the linear velocity of an object consists of the linear-velocity terms caused by the revolution, rotation, and precession. Apparently, the precessional motion is different from the revolution motion or rotation motion. And the necessary equation for precessional motion is distinct from that for the revolution motion or rotation motion.

The force and torque can generate the revolution motion and rotation motion respectively. The force is different from the torque. So the angular velocity of revolution, caused by the force, is considered to be independent of the angular velocity of rotation, caused by the torque. Both of which may be dissimilar in other physical properties either. For instance, these two types of angular velocities may consume (or store) different energies.

Similarly, the torque derivative is distinct from the torque. As a result, the angular velocity of precession induced by the torque derivative is absolutely different from that induced by the torque. They are two disparate types of angular velocities of precession. One obeys the octonion torque formula, Eq.(\ref{equ:3}), while the other meets the octonion force formula, Eq.(\ref{equ:4}). In terms of other physical properties, both of which may be diverse either.

\section{Experiment proposal}

In the complex-octonion space $\mathbb{O}$ , the influence of two constituents, $\textbf{g}$ and $\textbf{b}$ , of gravitational strength on the precessional angular velocity, $\overrightarrow{\omega}_p$ , is comparatively tough to be validated within the laboratory experiments at present. Only the influence of the constituent, $\textbf{b}$ , of gravitational strength may be observed in the astrophysical phenomena, especially the explanation of the astrophysical jets. However, it is able to validate directly the influence of the torque-derivative term, $v_0 \partial_0 \textbf{W}_1^i$ , and electromagnetic strength ($\textbf{E}$ and $\textbf{B}$) on the precessional angular velocity, $\overrightarrow{\omega}_p$ , in the laboratory. For instance, it is capable of measuring the influence of the torque-derivative term on the precessional angular velocity, for the rotating rotors. In the super-strong magnetic field (or electric field), it is able to explore the influencing degree of the external fields on the precessional angular velocities, in terms of the tiny charged objects.

a) Time-varying term. From Eq.(\ref{equ:18}), it is found that the precessional direction of the rotating rotor will orient the torque-derivative term, $v_0 \partial_0 \textbf{W}_1^i$, if the external torque, $- \textbf{W}_1^i$, is a time-varying term. In other words, when we accelerate or decelerate the rotation motion of rotors, it must generate the precessional motion. No matter the direction of torque-derivative term, $v_0 \partial_0 \textbf{W}_1^i$, is oriented or reversed to that of the rotation motion of rotors, it is still able to produce the precessional motion. Consequently, on the basis of existing gyroscopic experiments, it is feasible to validate the influence of the torque-derivative term on the precessional angular velocity in the experimental technique, after improving appropriately the existing experimental facilities.

b) Electromagnetic fields. According to Eq.(\ref{equ:20}), when there is the external magnetic field $\textbf{B}$ , the precessional direction of the rotating charged object will orient the vector, $\textbf{B} \circ \textbf{\emph{I}}_0$ . Even if the angular velocity of rotation is equal to zero, the external magnetic field is still able to induce the precessional angular velocity for the charged objects. Further, the spatial dimension of linear velocity will make a contribution to the precessional angular velocity of charged objects, especially the precessional motions in the three-dimensional magnetic fields. In the Electron Spin Resonance experiments to study the sample materials (such as, Di-Phehcryl Pircyl Hydrazal, CuSO$_4 \cdot$5H$_2$O, MnSO$_4 \cdot$4H$_2$O, or MnCl$_2 \cdot$4H$_2$O) with unpaired electrons, when we apply the Crystal Lattice Vibration methods (for instance, X-ray, $\gamma$-ray, or neutron inelastic scattering) to vibrate the crystal lattices, it is capable of generating the three-dimensional precessional motions of unpaired electrons surrounding the crystal lattices within the sample materials, measuring the absorption spectrum of electromagnetic waves caused by the three-dimensional precessional motions. Until now the Electron Spin Resonance experiment has never been validated under the furious vibration of crystal lattices. It is aggravating the existing serious qualms about the spin angular momentum. The paper appeals intensely to actualize the Electron Spin Resonance experiments in the drastic vibration of crystal lattices. It is predicted that the physical properties of spin angular momentum in this proposed experiment will be distinct from that in the conventional point of view.

In a similar way, from Eq.(\ref{equ:21}), when the external three-dimensional electric field $\textbf{E}$ is applied to the free and tiny charged objects (such as, hydrogen canal rays), we can also detect the spectrum of electromagnetic waves, caused by the two- or three-dimensional precessional motions. Certainly, the spectrum of electromagnetic waves in the external electric fields may be much more intricate than that in the external magnetic fields.

c) Gravitational fields. When there exists the external gravitational strength $\textbf{g}$, the precessional direction of the rotating neutral objects will orient the vector, $\textbf{g}$, from Eq.(\ref{equ:24}). No matter the direction of external gravitational strength $\textbf{g}$ , is oriented or reversed to that of the rotation motion of rotating neutral object, it is still able to induce the precessional motion. Even if the angular velocity of rotation is equal to zero, the external gravitational strength is still able to induce the precessional angular velocity for the neutral objects. Further, the spatial dimension of linear velocity will make a contribution to the precessional angular velocity of neutral objects, especially the precessional motions in the three-dimensional gravitational field, $\textbf{g}$ . Therefore, when we apply the external three-dimensional gravitational field, $\textbf{g}$ , to the tiny neutral objects, it is capable of measuring the precessional motions, although this precessional angular velocity may be difficult to detect.

The validation of these experiment proposals will be of great benefit to investigating the further features of precessional angular velocities.

\section{Conclusions and discussions}

The complex-quaternion spaces can be utilized to study the physical properties of either electromagnetic fields or gravitational fields. And the complex-octonion spaces can be applied to explore simultaneously the physical quantities of gravitational and electromagnetic fields, such as, the octonionic field strength, field source, linear momentum, torque, and force.

When the octonion force equals to zero, it is able to infer the eight independent equilibrium/continuity equations. Four of them are the force equilibrium equation, fluid continuity equation, precession equilibrium equation, and current continuity equation. The continuity equations belong to the equilibrium equations essentially. Moreover, the eight equilibrium/continuity equations are relevant to the spatial dimension of linear velocity.

From the precession equilibrium equation, it is found that the torque-derivative term, gravitational strength, and electromagnetic strength have an influence on the precessional angular velocities of neutral/charged objects. Especially, the magnetic flux density $\textbf{B}$ will make a contribution on the precessional angular velocities of charged objects, explaining the angular velocity of Larmor precession, for the spin angular momentum. And the gravitational precessional-angular-velocity $\textbf{b}$ is able to impact the precessional angular velocities of neutral objects, accounting for the precessional phenomena of astrophysical jets.

Either of torque and torque derivative may exert an influence on the angular velocity of precession. Especially, the external field is able to make a contribution on the precessional angular velocities of neutral/charged objects. Obviously, this precessional angular velocity is distinct from that caused by the conventional gyroscopic torque in the theoretical mechanics. The former obeys the octonion force formula, while the latter meets the requirement of the octonion torque.

It should be noted that the paper discussed only several simple cases about the influence of torque derivative on the angular velocities of precession. However it is clearly states that the external field strengths and the curl of linear velocity both are capable of inducing the precessional angular velocities of neutral/charged objects, including the magnitude and direction. In the following study, we will explore theoretically certain other influence factors of precessional angular velocities, by means of the precession equilibrium equation; and validate the influence of the curl of linear velocities on the precessional angular velocity in the three-dimensional magnetic fields, making use of some improved experiment devices; and research further the physical properties of spin angular momentum in the quantum mechanics, on the basis of the precessional angular velocity induced by the torque derivative.

\section*{Acknowledgements}

The author is indebted to the anonymous referees for their valuable comments on the previous manuscripts. This project was supported partially by the National Natural Science Foundation of China under grant number 60677039.

\appendix

\section{Gyroscopic torque}

In the complex-octonion spaces, the octonion torque formula can be degenerated into the torque formula (see Ref.\cite{weng1}). Later, in the complex-quaternion spaces, the torque formula is able to infer the precessional angular velocity caused by the conventional gyroscopic torque.

In the complex-octonion spaces, the octonion torque is Eq.(\ref{equ:3}), or
\begin{eqnarray}
\mathbb{W} = - v_0 ( \lozenge + i \mathbb{F} / v_0 ) \circ \mathbb{L}   ~,
\end{eqnarray}
further, in case there is no field strength, that is, $\mathbb{F} = 0$ , the above will be reduced into,
\begin{eqnarray}
\mathbb{W} = - v_0 \lozenge  \circ \mathbb{L}   ~.
\end{eqnarray}

In the complex-quaternion spaces, the above will be simplified into,
\begin{eqnarray}
\textbf{W}_1^i = - v_0 \partial_0 \textbf{L}    ~,
\end{eqnarray}
where $- \textbf{W}_1^i$ is conventional torque, which is the major ingredient of $\mathbb{W}$ . And $\textbf{L}$ is the orbital angular momentum, which is the primary constituent of $\mathbb{L}$ . The above is the definition of torque in the theoretical mechanics. Subsequently, according to the above, the conventional gyroscopic torque can produce the precessional angular velocity, which belongs to the torque-induced precessional angular velocity.

Apparently, the above is independent of the octonion force, Eq.(\ref{equ:4}), or
\begin{eqnarray}
\mathbb{N} = - ( \lozenge + i \mathbb{F} / v_0 ) \circ \mathbb{W}   ~.
\end{eqnarray}

According to the precession equilibrium equation, the torque derivative will induce another type of precessional angular velocity, which includes the field-induced precessional angular velocity. And this is distinct from the precessional angular velocity caused by the gyroscopic torque.

\end{document}